\documentclass[12pt]{iopart}

\usepackage{graphicx}
\graphicspath{ {./images/} }
\usepackage{hyperref}

\usepackage{xcolor}
\usepackage{amsmath}

\begin{document}

\title[]{Measurements of the amplitude-dependent microwave surface resistance of an Au/Nb bilayer}

\author{Thomas Oseroff, Zeming Sun, and Matthias U. Liepe}

\address{Cornell Laboratory for Accelerator-Based Sciences and Education, Cornell University, Ithaca, NY 14853, USA}
\ead{teo26@cornell.edu}
\vspace{10pt}
\begin{indented}
\item[May 2023]
\end{indented}

\begin{abstract}
Surface properties are critical to the capabilities of superconducting microwave devices.  The native oxide of niobium-based devices is thought to consist of a thin normal conducting layer.  To improve understanding on the importance of this layer, an attempt was made to replace it with a more easily controlled gold film.  A niobium sample host microwave cavity was used to measure the surface resistance in continuous wave operation at $4.0\,\text{GHz}$ and $5.2\,\text{GHz}$.   Sample conditions studied include temperatures ranging from $1.6\,\text{K}$ to $4.2\,\text{K}$ with RF magnetic fields on the sample surface ranging from $\sim 1\,\text{mT}$ to the maximum field before the superconducting properties were lost (quench field).  The nominal film thickness of the gold layer was increased from $0.1\,\text{nm}$ to $2.0\,\text{nm}$ in five steps to study the impact of the normal layer thickness on surface resistance on a single niobium substrate.  The $0.1\,\text{nm}$ film was found to reduce the surface resistance of the sample and to enhance the quench field.  With the exception of the final step from a $1.5\,\text{nm}$ gold film to $2.0\,\text{nm}$, the magnitude of the surface resistance increased substantially with gold film thickness.  The nature of the surface resistance field-dependence appeared to be roughly independent from the gold layer thickness.  This initial study provides new perspectives and suggests avenues for optimizing and designing surfaces for resonant cavities in particle accelerators and quantum information applications.

\end{abstract}

%
\vspace{2pc}
\noindent{\it Keywords}: SRF, superconducting radio frequency cavities, proximity-coupling, surface resistance, niobium

%
\submitto{Superconductor Science and Technology}
%
\maketitle
%
\ioptwocol

\section{Introduction}

The properties of niobium make it desirable for a wide range of superconducting microwave applications~\cite{Hasan_50_years}.  These include low-field devices such as detectors and elements of quantum circuits, as well as high-field resonant cavities used in particle accelerators.  Independently of the specific purpose or conditions of device operation, its optimization often depends on precise control of the surface properties where the microwave-superconductor interaction is strongest.

Due to the importance of surface features and properties in microwave superconducting devices, the oxidation of the surface, present on most niobium devices, is expected to be relevant.  
The native niobium oxide is usually less than $10\,\text{nm}$ thick, and roughly comprises three layers of different oxidation states~\cite{DESY_oxide}.  The outermost layer is $\text{Nb}_2\text{O}_5$, behaving as a dielectric.  Two-level systems in this layer act as a dominant loss mechanism for low-field and low-temperature devices~\cite{FNAL_TLS_2017,FNAL_TLS_2020,FNAL_TLS_2022}.  The layer closest to niobium is subnanometer metallic $\text{NbO}_x$ ($x\leq1$).  At the higher temperatures ($> 1.5\,\text{K}$) used for high-field device operation, this metallic oxide is a normal conductor.  In addition to dissipation occurring directly in the oxide, the normal conducting portion will also influence the properties of the superconducting niobium surface through proximity-coupling~\cite{Gurevich_SN}.  Understanding the underlying role of a thin normal-conducting layer on the relevant metrics is essential for optimizing resonant cavities used in particle accelerators and providing routes of surface design for other microwave superconducting devices.

A model of the surface resistance in a dirty proximity-coupling system predicts low-field values that can be less than or greater than that expected for a bare superconductor, depending on the normal conductor properties, superconductor properties, and their electrical connection~\cite{Gurevich_SN}.  An extension of this model to include high-field pair-breaking effects predicts that these proximity-coupling properties can also be significant in determining surface resistance field-dependence~\cite{Kubo_SN}.  The calculation of density of states used in these models has been compared to relevant data~\cite{Temple_tunneling}, but the estimates of surface resistance have not been experimentally considered.

Directly measuring the impact of the metallic oxide is challenging, as it is complicated to control the properties of the niobium oxide layer at nanoscales in a microwave cavity that is capable of high-field operation.  Attempts have been made to study the effect of the oxide through high-temperature \emph{in situ} baking on an assembled cavity measurement apparatus under vacuum~\cite{Eremeev_thesis}.  By controlling the subsequent oxygen exposure, the oxide structures and thicknesses can be modified.  The exact effect this has on the properties of the oxide is difficult to understand due to its structural complexity and high reactivity~\cite{Zeming_oxide}.  The measurement is further obscured by other effects induced by the high-temperature treatment, such as impurity diffusion.  The results of their study indicate that altering the oxide could be relevant for improving accelerator cavities, but further conclusions are challenging due to the lack of more precise control of relevant properties.

In the present work, an attempt was made to artificially replace the complex niobium oxide with a passivating gold film, and to then study the surface resistance of the Au/Nb system with different gold layer thicknesses.  Because gold is inert and will not react with oxygen, there will be no complications resulting from a new oxide layer.  The relevant properties, such as the gold layer thickness, conductivity, and its contact resistance to the niobium substrate, are more easily understood and controlled than those of the niobium oxide.  The aim is to provide the first experimental results on the surface resistance field-dependence of a controlled normal conductor / superconductor system.


Measurements of the low-field microwave surface resistance of proximity-coupled systems have been reported previously~\cite{Pan_Au_YBCO,Pambianchi_Al,Pambianchi_Cu,Pambianchi_thesis}.  The present study extends the results of Pambianchi and Anlage on sputtered Al/Nb and Cu/Nb films to consider the high-amplitude field dependence of the surface resistance.  To better explore the behavior for strong RF fields, bulk niobium substrates are considered instead of a sputtered film that may encounter unnecessary uncertainties from structure distortion, strain, and impurity incorporation during deposition~\cite{AMVF_big_paper}.


The paper is organized as follows.  The sample host cavity used to produce large fields on the Au/Nb system and measure its surface resistance is introduced in section \ref{sec:BOB_cavity}.  The preparation procedure for the Au/Nb bilayer is detailed in section \ref{sec:sample_prep}.  The results of the study are presented in sections \ref{sec:Thin_gold} and \ref{sec:results_thickness}.  

The results indicate that controlling or replacing the niobium oxide layer could be a path to reduced surface resistance in high-field applications.  This enhancement could be relevant for low-field applications as well.  Guided by models suggesting that proximity-coupling can reduce total surface resistance at low fields~\cite{Gurevich_SN}, it is not unreasonable to assume that the observed low surface resistances would persist at very small field amplitudes.  If this speculation holds, then a similar method for passivating a niobium surface using a gold layer to that which was considered here may illuminate a viable path for eliminating the dissipation caused by two-level systems in the dielectric portion of the niobium oxide.  This could be used to improve coherence times of quantum states coupled to superconducting resonators~\cite{FNAL_TLS_2017,FNAL_TLS_2020,FNAL_TLS_2022}.

\section{Sample excitation and measurement}
\label{sec:BOB_cavity}


This section describes the resonator used to excite large microwave fields on the sample, the method used to extract the sample surface resistance, and the magnetic conditions present on the sample when it transitions into the superconducting state.  This information is essential for correctly interpreting key results in section \ref{sec:Thin_gold}.

\subsection{Sample host cavity}

\begin{figure*}[!htbp]
    \centering
    \includegraphics[width = \linewidth]{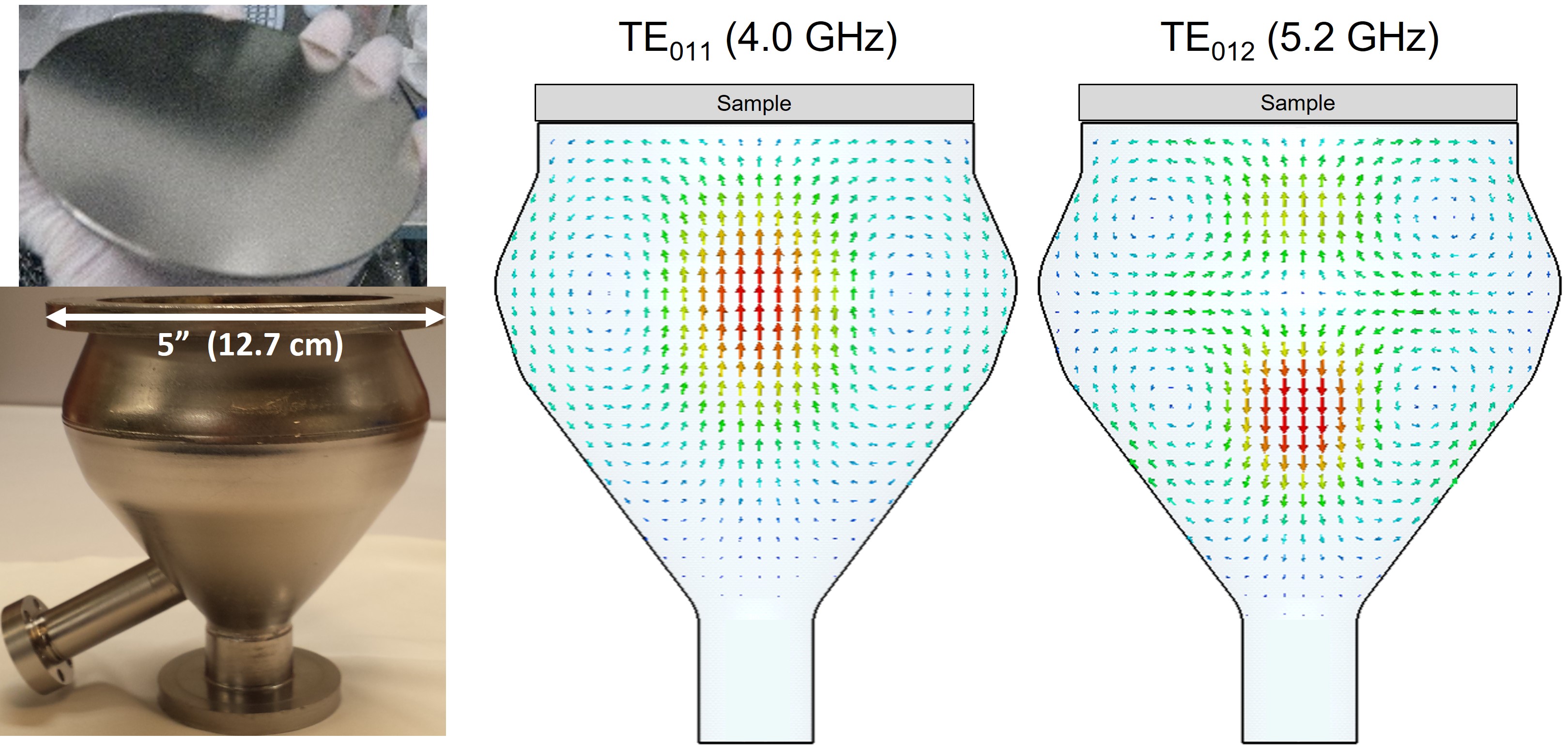}
    \caption{(Left) Sample host cavity (bottom) and sample disk/plate (top).  The marked $5\,\text{inch}$ ($12.7\,\text{cm}$) opening marks the location where the sample joins the host structure.  (Right) CST Microwave Studio simulation results showing the cross-sectional magnetic fields of the operating modes.}
    \label{fig:BOB_cavity}
\end{figure*}

The samples were exposed to strong RF fields using a sample host microwave cavity designed to operate with $\text{TE}_{011}$-like and $\text{TE}_{012}$-like modes at $4\,\text{GHz}$ and $5.2\,\text{GHz}$ respectively~\cite{Hall_SRF_2013,Oseroff_thesis}.  The $12.7\,\text{cm}$ diameter $3\,\text{mm}$ thick sample disk is affixed to the opening at the top of the host structure to close the volume and complete the resonant structure demonstrated in figure \ref{fig:BOB_cavity}.  The magnetic field profile at the surface of the sample disk is spatially inhomogeneous, but similar for both operating modes.  For the remainder of the text, the "sample field" will refer to its value at the location of the maximum.

The host structure geometry was engineered to maximize the obtainable magnetic field amplitude on the sample before being limited by the niobium host structure~\cite{Oseroff_thesis}.  In addition to the opening for the sample, two ports were included for coupling power into and out of the resonator.  Simulated cavity metrics are given for each operating mode in table \ref{tab:BOB_params}.  

The host structure was constructed from fine-grain niobium~\cite{Hall_SRF_2013}.  The formed cavity was electropolished (EP) to remove over $100\,\mu\text{m}$ of material to eliminate unwanted inclusions from the manufacturing processes and to produce a smooth surface to reduce field enhancement~\cite{Oseroff_thesis}.  This was followed with an $800^\circ\,\text{C}$ bake in ultra-high vacuum ($10^{-10}\,\text{Torr}$) for five hours to remove hydrogen introduced by the EP acids.  A final $4\,\mu\text{m}$ EP was conducted to remove any surface impurities introduced from the high-temperature bake.  A high-pressure rinse using deionized (DI) water was used prior to the furnace treatment and before any RF measurements to remove contamination.  This preparation ensures low surface resistance over a wide range of RF field amplitudes for which the superconducting state can be maintained.  

\begin{table}[!htbp]
    \centering
    \caption{Cornell sample host cavity parameters calculated from results of a CST Microwave Studio simulation.  $B$ refers to the magnetic field.  The subscript "$pk$" denotes the peak value on the specified surface.  $\alpha = \frac{ \int_{plate} |B|^2 dS }{\int_{plate} |B|^2 dS + \int_{host} |B|^2 dS  }$ is referred to as the focusing factor and indicates the fraction of surface field magnitude seen by the sample.  $U$ is the stored energy in the resonator.  $G$ is the cavity geometry factor \cite{Pozar_book}.  The $\text{TE}_{012}$ mode deviates slightly from simulation and is measured to be 5.23 GHz.}
    \begin{tabular}{l|r|r}
        & $\text{TE}_{011}$ & $\text{TE}_{012}$ \\\hline
        frequency [GHz] & 3.99 & 5.27 \\\hline
        $\frac{B_{pk,plate}}{B_{pk,host}}$ & 0.91 &  0.96\\\hline
        $\alpha$ & 0.13 & 0.25\\\hline
        $\frac{B_{pk,host}}{\sqrt{U}}\,\left[\frac{\text{mT}}{\sqrt{\text{J}}}\right]$ & 69.3   & 94.2 \\\hline
        $G\,\left[\Omega\right]$& 801& 993
    \end{tabular}
    \label{tab:BOB_params}
\end{table}

\subsection{Quality factor measurement}

The quality factor of the resonator is measured by driving the cavity to a steady-state at resonance using a phase-locked loop and then turning off the RF drive.  The subsequent resonator energy decay is monitored through the RF power leaking from the cavity.  The quality factor is obtained by fitting this decay to obtain a characteristic time of an assumed exponential form.  The domain of the fit spans from the first point after the RF drive is shut off to the time required for the signal to be reduced to $81\%$ of its initial value.  Combining the characteristic decay time with the power-coupling information obtained from transmission line analysis and steady-state power measurements, the intrinsic quality factor of the cavity can be obtained~\cite{Oseroff_thesis}.  

\subsection{Extracting the sample surface resistance}

To extract the surface resistance of the sample, its contribution to the measured quality factor must be isolated.  The method employed here for this purpose will be referred to as a calibrated quality factor measurement, where two separate measurements are used to remove the contribution of the host structure to the measured quality factor.  First, a calibration measurement is performed, characterized by using a calibration sample prepared identically to the host structure.  It is assumed that the surface resistance of the host structure and this calibration sample are equal.  Second, the quality factor of the resonator assembled with the sample plate of interest is measured.  It is assumed that the dissipation on the host structure is the same for the calibration measurement and the sample measurement, and that additional systematic contributions to the quality factor are minimal.  With the preceding assumptions, the surface resistance of the sample, $R$, is related to the measured quality factors 
\begin{equation}
    R = \frac{G}{\alpha} \left[ \frac{1}{Q_0^{sam}} - (1-\alpha)\frac{1}{Q_0^{cal}} \right].
    \label{eqn:Sample_R_extraction}
\end{equation}
$G$ and $\alpha$ are the cavity geometry factor and focusing factor described in table \ref{tab:BOB_params}.  $Q_0^{sam}$ and $Q_0^{cal}$ are the intrinsic quality factors measured with the sample plate and the calibration plate respectively.

The calibrated quality factor method to extract sample surface resistance is most effective when the sample surface resistance is large compared to that of the host structure.  The method becomes increasingly susceptible to uncertainty as the sample surface resistance is reduced below that of the host structure.  In particular, the systematic uncertainties introduced by inadequate control of the previously described assumptions can be problematic in this limit.

Steps are taken throughout the process to mitigate such issues.  The host structure and sample plate are cleaned using a high-pressure rinse with DI water prior to assembly in a class 100 cleanroom environment to prevent surface contamination that could lead to variation between measurements.  To minimize changes in dissipation due to trapped magnetic flux~\cite{Gurevich_vortex,Danilo_vortex,Checchin_vortex}, the helium transfer is conducted in a magnetically shielded cryostat and the ambient magnetic fields are monitored.  Despite these precautions, systematic errors are expected to be relevant, especially when the sample surface resistance is small compared to that of the host structure.

\subsection{RF field amplitude measurement}

Measurement of the field at the sample surface is obtained from forward and reverse power measurements~\cite{Oseroff_thesis}.  From this and the measured quality factor, the energy stored in the resonator can be obtained and then converted to the surface field magnitude using parameters from table \ref{tab:BOB_params}.  Sources of error are present in this measurement, but it is independent of the calibration procedure.  Therefore, the RF field values reported in sections \ref{sec:Thin_gold} and \ref{sec:results_thickness} are as reliable as typical cavity measurements~\cite{FNAL_cavity_test_errors}.

\begin{figure*}[!htbp]
    \centering
    \includegraphics[width = \linewidth]{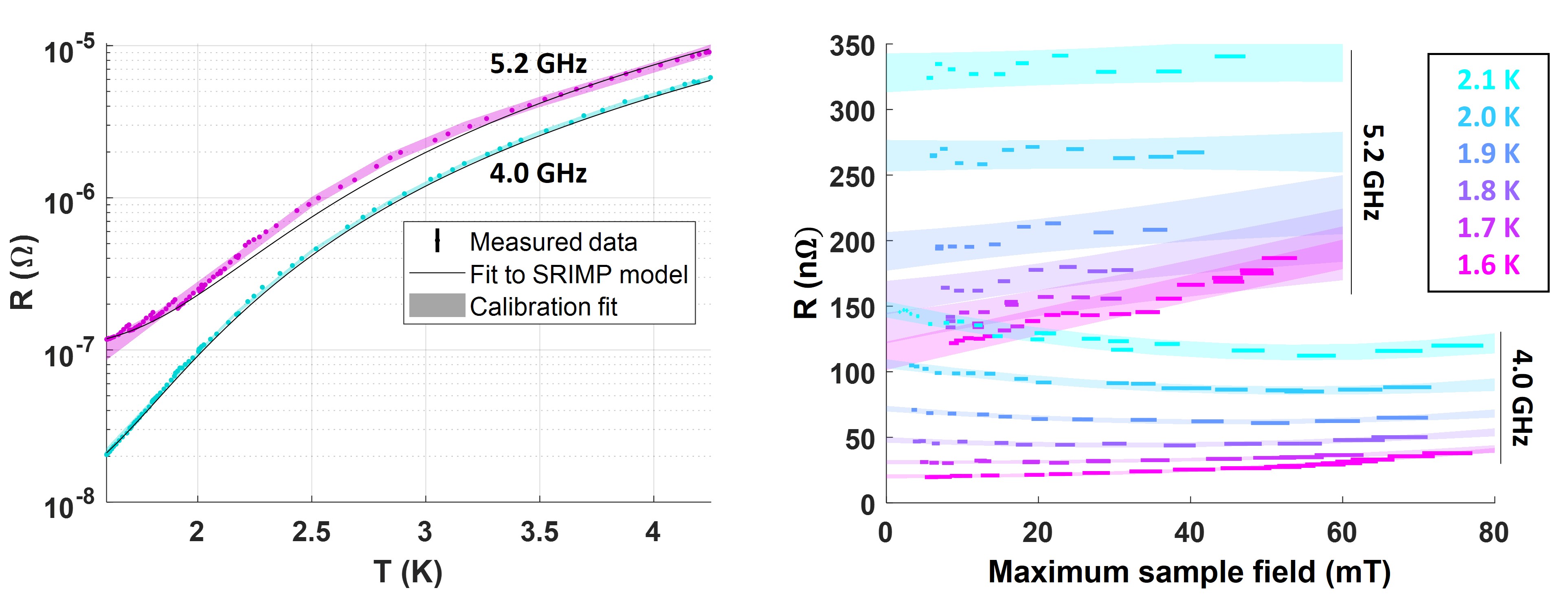}
    \caption{Calibration (niobium) measurement of average surface resistance, $R = G/Q_0$.  Low field (see figure \ref{fig:RvsT} for amplitudes) measurements were taken for a range of temperatures (left).  Data was fit using the SRIMP model~\cite{halbritter_SRIMP} to demonstrate agreement with theoretical expectations.  Input power was varied at select temperatures (right).  The shaded regions correspond to polynomial fits used for calibrating sample data at a specific temperature or field strength.}
    \label{fig:BOB_calib}
\end{figure*}

\subsection{Calibration measurement}

The calibration measurement results are demonstrated in figure \ref{fig:BOB_calib}.  This measurement, and all others presented later, were performed with the cavity completely immersed in liquid helium.  The cryostat pressure was controlled to vary the temperature.  The surface resistance was measured with small input power as the temperature was slowly decreased at a rate of approximately $1\,\text{K/hr}$.  The resulting values are fit to predictions of the surface impedance model, "surface resistance with impurities" (SRIMP)~\cite{halbritter_SRIMP}, describing dissipation due to quasiparticle excitations with properties corresponding to those in the interior of a bulk superconductor.  The best-fit parameter values agree well with those expected of niobium~\cite{Oseroff_thesis}.

On the right side of figure \ref{fig:BOB_calib}, surface resistance was measured at constant temperatures for a range of input powers.  The helium bath temperature was held constant within several millikelvin at values ranging from $2.1\,\text{K}$ to $1.6\,\text{K}$ in steps of $0.1\,\text{K}$.  The input power was varied at each temperature to explore the effect of field amplitude on the electromagnetic response.  Published data exploring niobium at high RF field amplitudes in this frequency range is limited, and is only available at $4.0\,\text{GHz}$~\cite{Fermi_high_frequency}.  Despite being excited with a different electromagnetic mode, the results here agree well with those of the Fermilab study and include the reported anomalous field effect ("Anti-Q-slope")~\cite{Oseroff_thesis}.

The shaded regions in figure \ref{fig:BOB_calib} correspond to polynomial fits to the data with a $95\%$ confidence interval.  The purpose of this fit is for calibration of sample measurements taken at temperatures/fields between those of the calibration data.  The polynomial order was chosen somewhat arbitrarily to be fourth order for the low-amplitude $R(T)$ data, second order for the $4.0\,\text{GHz}$ $R(B)$ data, and first order for the $5.2\,\text{GHz}$ $R(B)$ data.  As is evident in figure \ref{fig:BOB_calib}, the linear fit of the $5.2\,\text{GHz}$ field-dependence does not capture its finer detail.  This detail was observed to vary when measurements were repeated.  The choice to use a linear fit was made to effectively remove this variation from the calibration, at the cost of increased uncertainty.

The confidence interval limits of the fit are used to convey uncertainty in the calibration measurement.  The statistical uncertainty results mostly from noise on the power meters and the previously described acquisition of the characteristic decay time, and is reported to be very low.  This calibration measurement uncertainty will propagate into the reported sample surface resistance through equation \ref{eqn:Sample_R_extraction}.  

\subsection{Magnetic conditions of the sample}

When niobium is cooled to the superconducting state in the presence of a magnetic field, magnetic flux vortices can be trapped by impurities and defects.  These trapped vortices can act as an additional source of dissipation in an RF field~\cite{Gurevich_vortex,Danilo_vortex,Checchin_vortex}.  Because of this, it is important to consider the magnetic conditions in the cryostat. 

Ideally the magnetic fields will be minimal and consistent between related measurements.  Ambient magnetic fields are reduced by magnetic shielding to be less than $10\,\text{mG}$ near the sample host cavity.  This value was reasonably consistent between the calibration measurement and those with the gold layers to be discussed in sections \ref{sec:Thin_gold} and \ref{sec:results_thickness}.  From this, it can be argued that ambient magnetic fields are not acting as a significant source of systematic error in the calibration procedure or as a source of extrinsic variation between sample measurements.


It is thought that larger thermal gradients are more effective for expelling magnetic flux past trapping centers~\cite{FNAL_thermal_gradient_cooling,Kubo_thermal_gradient}.  However, thermal currents produced by thermal gradients along layered metallic structures due to the Seebeck effect can produce significant magnetic fields~\cite{Posen_and_Hall_2017}.  This problem has been studied experimentally and modeled for niobium cavities with films of $\text{Nb}_3\text{Sn}$~\cite{Hall_thesis}.  Applying this model to the Au/Nb structures considered in the remaining sections, it was found that the gold layers are too thin to generate significant magnetic fields from this effect.  Therefore, it was decided that trapped flux dissipation would be minimized by quickly transferring helium to the cryostat to maximize the thermal gradients.  The thermal gradient from the bottom of the cavity to the sample plate was approximately $15\,\text{K}$ for all measurements discussed in sections \ref{sec:Thin_gold} and \ref{sec:results_thickness}.  


\section{Sample preparation}
\label{sec:sample_prep}



A niobium substrate was prepared by attempting to remove its native oxide and then replace it with an inert gold layer.  The gold layer pacifies the surface preventing further oxidization.  The thickness of this gold layer was incrementally increased.  Between each addition to the gold layer thickness, an RF measurement was carried out to measure how the normal conductor thickness impacts surface resistance in the Au/Nb bilayer.

Two separate niobium plates are considered in this work; the calibration plate and a plate for the gold layers.  The niobium substrate used for the gold layer study was prepared identically to the calibration plate described in section \ref{sec:BOB_cavity}.  It was then measured with similar conditions to the calibration data in figure \ref{fig:BOB_calib}.  This measurement provides a baseline for performance prior to creation of the Au/Nb bilayer.  At this stage, the two plates are functionally identical and produce similar results.

To remove the native oxide layer, the sample was loaded into a nitrogen gas-filled glove box with $\text{O}_2$ and $\text{H}_2\text{O}$ levels below $0.5\,\text{ppm}$.  The sample was soaked in diluted hydrofluoric acid (HF) ($1-2\,\%$ in DI water) for 30 minutes to etch the surface oxide.  Chemical etching is favored over typical baking approaches for oxide modification since heat treatments at less than $2200^\circ\,\text{C}$ are unable to remove the metallic portion of the oxide that is relevant for this study~\cite{CBB_Chicago_oxide}.  Glovebox operation also allows for \emph{in situ} sealings to avoid post-processing air exposure.  Methanol was used to remove residual HF and $\text{H}_2\text{O}$ from the sample surface.  It was then left to dry in the nitrogen atmosphere before being placed into three individually  sealed plastic bags for transport to a gold deposition system.  X-ray photoelectron spectroscopy on the fabricated Nb surface (with a capping layer) confirms the elimination of any oxygen at the Nb surface.

The gold layer was deposited using a CVC SC4500 electron-beam evaporation system.  To minimize/prevent oxidation during sample loading, the sealed bag was placed in the chamber with the nitrogen purge active and allowed to flow for some time before pumping the system to less than $10^{-6}\,\text{Torr}$.  

The gold deposition rate used was $0.01\,\text{nm/s}$.  The final thickness was monitored by calibrated Quartz Crystal Control.  Prior to deposition onto the sample, the ingot was cleaned by pre-evaporating $15\,\text{nm}$ of gold.

The first step was depositing a nominal $0.1\,\text{nm}$ layer of gold and performing RF measurements.  After this, more gold thickness was added and measured with the sample host cavity.  This process was repeated four times, achieving a maximum  nominal gold thickness of $2.0\,\text{nm}$.  RF measurement with the sample host cavity requires crushing an indium gasket along the perimeter of the sample.  After each measurement, this was removed with a copper edge and methanol.  To mitigate contamination, the sample was treated with a DI water high-pressure rinse before the deposition of more gold, and then again prior to re-assembly onto the host structure.

To validate this method of reducing/eliminating niobium oxidation, contact resistivity (resistance between layers multiplied by contact area) measurements were carried out on separate niobium samples.  The gold was patterned on these samples through a mask to create individual contacts on the niobium surface.  One sample was prepared according to the procedure just described.  The other was prepared without attempting to minimize niobium oxidation.  The resistance between layers was obtained using a four-probe measurement using the AC transport mode of a Quantum Design DynaCool physical properties measurement system (PPMS).  The contact resistance was isolated by measuring the total resistance below the niobium critical temperature to eliminate the contribution from the substrate.  Because the $50\,\text{nm}$ gold layers were too thin to contribute significantly to the total resistance, the only remaining source was the contact resistance between the gold and niobium.  The contact resistivity measured for the sample with minimal oxide was $6\times 10^{-13}\,\Omega\cdot\text{m}^2$, which was an order of magnitude less than that of the sample with standard niobium oxide.  The reduction in contact resistance indicates that the surface passivation method was, at least to some degree, successful.

The contact resistance measurement shows that gold is being deposited prior to complete oxidation.  The initial gold deposition for the sample used in the RF measurements of sections \ref{sec:Thin_gold} and \ref{sec:results_thickness} was much less than that in the contact resistance measurement.  This may not provide complete coverage of the surface.  If such a scenario occurs, portions of the surface could be oxidized while others have bonded with gold.  All results should be interpreted as having an unknown fraction of the surface being oxidized.  Even if only a fraction of the surface has changed, it is expected that the total dissipation on the surface will shift and lead to qualitatively meaningful data.

\section{Effects of attempting to reduce the native oxide}

\label{sec:Thin_gold}

 \begin{figure*}[!htbp]
    \centering
    \includegraphics[width = \linewidth]{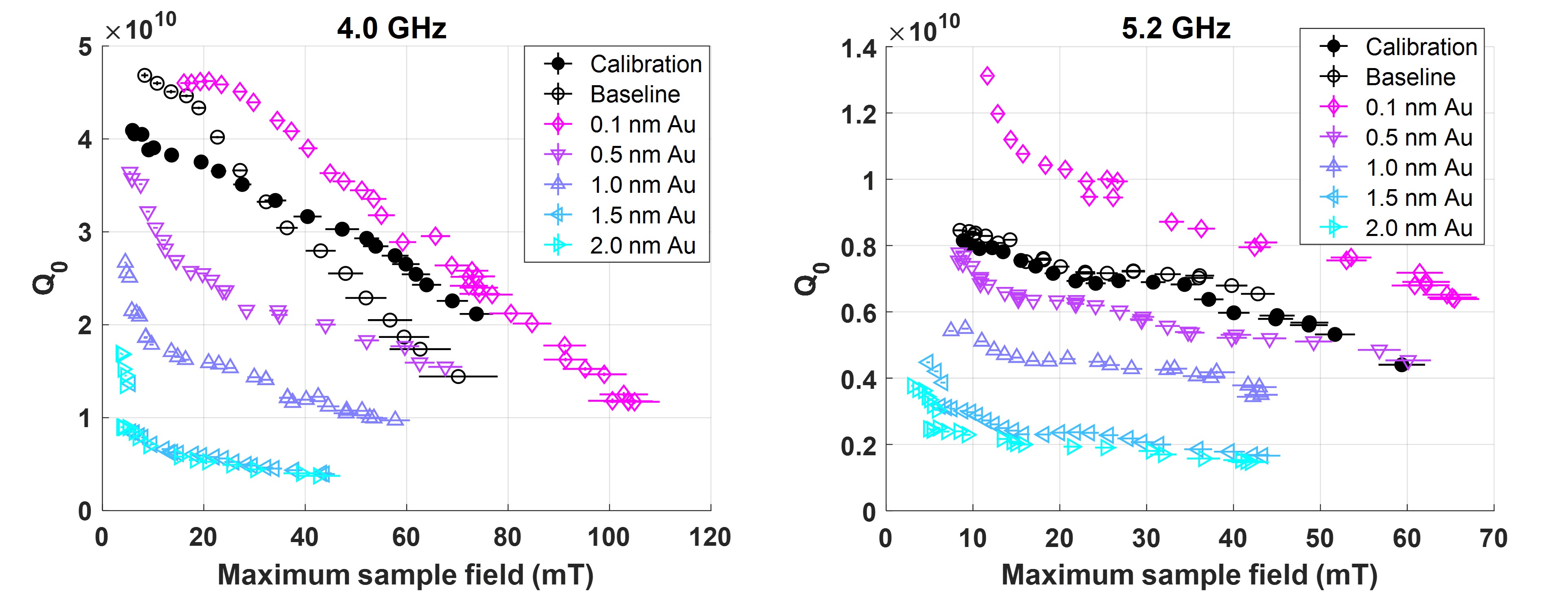}
    \caption{Intrinsic quality factor measured at 1.6 K and $4.0\,\text{GHz}$ (left) and $5.2\,\text{GHz}$ (right).  The baseline and calibration measurements are separate naturally oxidized niobium substrates prepared as described in section \ref{sec:BOB_cavity}.  The gold film measurements, prepared by removing the oxide layer of the "baseline" substrate as described in section \ref{sec:sample_prep}, were performed after each addition to the gold thickness.  The host structure surface was reset between the baseline measurement and those where its native oxide was replaced with gold, so some of the change in quality factor between the baseline and gold measurements could be attributed to the host structure.  Maximum fields were limited by quench (the maximum field amplitude that can support superconducting properties).}
    \label{fig:Q0_vs_B}
\end{figure*}

An attempt was made to observe the effect a niobium oxide layer has on surface resistance and its dependence on RF field strength by comparing the measurements of niobium plates with the initial oxide present to one where it is presumed to be replaced by the thinnest gold layer (nominally $0.1\,\text{nm}$, but for brevity, the specification that this thickness is in-name-only will be omitted).  This minimal gold thickness ideally will approximate a bare niobium surface, as it should have a negligible effect on the surface resistance while preventing oxidation.  The calibrated quality factor measurement, described in section \ref{sec:BOB_cavity}, is ineffective for extracting the surface resistance in this case, as it is limited when sample surface resistance becomes smaller than that of the host structure.

 While it is not possible to obtain meaningful surface resistance values for the $0.1\,\text{nm}$ gold sample, it is valid to compare the measured quality factors with the $0.1\,\text{nm}$ gold sample to those with the native oxide, which is done in figure \ref{fig:Q0_vs_B}.  This section will focus on comparing the oxidized niobium of the baseline and calibration measurements to those of the $0.1\,\text{nm}$ gold layer sample.  The subsequently measured thicker gold layers are included in figure \ref{fig:Q0_vs_B}, but will not be discussed until section \ref{sec:results_thickness}.

Comparing the quality factors of the resonator with two different sample plate surfaces conveys relative shifts in the surface resistance of the sample plates.  A given sample having a higher (lower) quality factor indicates a reduced (increased) surface resistance as compared to that of another sample.  In addition to this qualitative information, figure \ref{fig:Q0_vs_B} can be used to quantitatively compare the quench field associated with each sample.  The quench field is defined as the maximum field amplitude that can support superconducting properties.  The point of highest field for each specified measurement in figure \ref{fig:Q0_vs_B} corresponds to the lowest input power for which a quench event was observed attenuated by $1\,\text{dB}$.

When considering the quality factor of the baseline sample, the reader should be aware that the host structure surface had to be reset (preparation described in section \ref{sec:BOB_cavity}) after the baseline measurement due to a technical issue.  The changes between the baseline measurement and the other measurements considered could be, in part, due to slight differences in the host structure.  The host structure surface was unchanged between the other measurements presented in figure \ref{fig:Q0_vs_B}.

The quality factor measurements with the $0.1\,\text{nm}$ gold layer are higher than those of both the calibration and baseline plates.  Without the influence of significant systematic error, this suggests that replacing or reducing the native oxide with a very thin gold film reduced the surface resistance and increased the quench field.  The likelihood that the increased quality factor and quench field are caused by systematic error is now considered.


The increase in quality factor is not expected to be due to changes in the surface resistance of the host structure.  As discussed in section \ref{sec:BOB_cavity}, trapped flux dissipation in the host structure is expected to be similar in all measurements.  The magnetic conditions in the cryostat during the helium transfer were monitored and displayed no unusual values for the $0.1\,\text{nm}$ gold layer measurement.  Moreover, the improved quality factor is observed at both measurement frequencies and the cavity is thermally cycled to room temperature between the $4.0\,\text{GHz}$ and $5.2\,\text{GHz}$ measurements.  Contamination on the host structure is unlikely to be relevant due to the care taken in the process of changing sample plates, described in section \ref{sec:BOB_cavity}.

The increase in quench field is not expected to be due to issues in the RF path.  Path attenuations and cavity coupling factors are measured separately for each sample preparation and displayed no unusual values or behaviors with the $0.1\,\text{nm}$ gold layer.  The increased quench field is observed at both measurement frequencies.  From this, it is unlikely that a bad attenuation measurement or component could account for the observed increases because many elements of the RF pathways are changed for measurement at the different frequencies and the RF path attenuations are measured separately for each frequency.

While monitored systematic effects are not expected to have created the enhanced quality factors and quench fields in the $0.1\,\text{nm}$ gold film measurement, the magnitude of the increases in both metrics over the values of the calibration measurement is too large to be explained without the influence of systematic error.  To explain these values, the validity of the core assumptions for the calibrated quality factor measurement, described in section \ref{sec:BOB_cavity}, must be examined.  

If the assumptions leading to equation \ref{eqn:Sample_R_extraction} are satisfied, the maximum quality factor would be $Q_0 = Q_0^{cal}/(1-\alpha)$.  This corresponds to the case of negligible sample surface resistance, $R = 0$.  From the focusing factors specified for each mode in table \ref{tab:BOB_params}, it is observed that the measured quality factors with the $0.1\,\text{nm}$ sample exceed this maximum value.

The increase in quench fields is also inharmonious with the assumptions leading to equation \ref{eqn:Sample_R_extraction}.  From table \ref{tab:BOB_params}, it is observed that the peak fields on the plate should be slightly lower than those on the host structure in both modes.  Combined with the assumption that the surface resistance of the host structure and calibration plate are identical, the quench field in the calibration measurement should occur on the host structure.  With the assumption that the host structure is unchanged between measurements, one can conclude that obtainable quench fields cannot exceed those of the calibration measurement.


As mentioned previously in this section, the assumption that the host structure surface resistance is unchanged between measurements appears to be feasible.  More problematic is the assumption that the calibration plate surface resistance is identical to that of the host structure.  It has been observed that the sample plates are more susceptible to contamination introduced in the high temperature bake described in section \ref{sec:BOB_cavity}~\cite{Oseroff_thesis}.  A light electropolish, consistently reduces the measured surface resistance of calibration measurements to values consistent with expectation from theory and experiment (as discussed with figure \ref{fig:BOB_calib}).  However, the same quality factor measurement could be obtained if the calibration plate surface resistance was increased, due to lingering contamination from the furnace, while that of the host structure was slightly decreased.

This scenario, that the calibration plate has a slightly larger surface resistance than the host structure, can explain the issues with the quality factors and quench fields in the $0.1\,\text{nm}$ gold layer measurement.  It is assumed that the quench mechanism is a thermal breakdown, where the superfluid helium is unable to remove heat from the system efficiently enough to maintain the superconducting state~\cite{Gurevich_thermal_breakdown}.  At the measurement frequencies considered here, this is likely the case~\cite{yi_xie_thesis}.  A simple thermal analysis~\cite{Gurevich_thermal_breakdown} using reasonable thermal properties for niobium~\cite{Ryan_thesis} indicates thermal breakdown fields consistent with all measurements in figure \ref{fig:Q0_vs_B}.  The surface resistances required to produce the observed quench fields with a thermal breakdown mechanism are consistent with the surface resistance of the calibration plate being larger than that of the host structure.  The increase in calibration plate surface resistance is sufficient to justify the quench occurring on the plate instead of the host structure, which provides a path to explain the increased quench field in the $0.1\,\text{nm}$ gold layer measurement.  To produce the observed quality factor, the increase in calibration plate surface resistance requires a slight reduction to the surface resistance on the host structure compared to the assumed $G/Q_0$.  With this reduction, the magnitude of the observed quality factors of the $0.1\,\text{nm}$ gold layer sample can be understood.

The goal of this section is to help clarify the influence of niobium oxide on microwave surface resistance.  It can be reasonably concluded that the surface resistance of the niobium plate was reduced following the attempted passivation procedure described in section \ref{sec:sample_prep}.  While this procedure is imperfect, it likely reduces the oxide, even if not fully replacing it with a $0.1\,\text{nm}$ gold layer, as was discussed in section \ref{sec:sample_prep}.  It can be argued that the niobium oxide had a detectable and detrimental effect on the surface resistance of the sample plate by combining the arguments in section \ref{sec:sample_prep} for some fraction of the surface having reduced oxide, intuition that a thinner normal conducting layer can lower surface resistance~\cite{Gurevich_SN,Kubo_SN}, and the observed reduction in sample surface resistance.  However, due to the necessity of including systematic error to explain the measurements, it cannot be concluded that this result extends beyond this sample.  These measurements cannot rule out the case that the sample surface resistance was initially high compared to “typical” niobium with the same preparation.  It is clear that the preparation procedure reduced the surface resistance of the sample, but it could still be higher than “typical” oxidized niobium.

\begin{figure*}[!htbp]
    \centering
    \includegraphics[width = \linewidth]{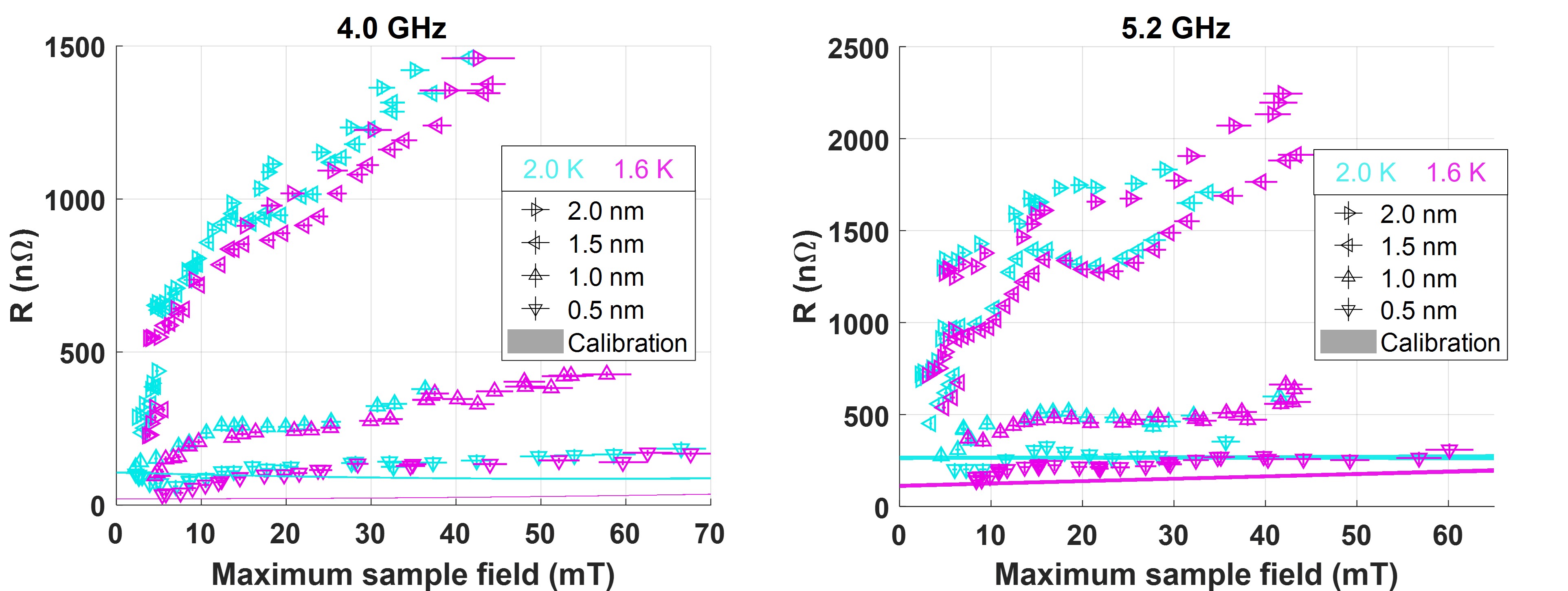}
    \caption{Extracted sample surface resistance (equation \ref{eqn:Sample_R_extraction}) of thin gold layers deposited over bulk niobium at $1.6\,\text{K}$ (cyan) and $2.0\,\text{K}$ (pink) driven by RF fields at  $4.0\,\text{GHz}$ (left) and $5.2\,\text{GHz}$ (right). All measurements were performed on a single niobium sample plate with a gold layer replacing the native niobium oxide as described in section \ref{sec:sample_prep}.  The calibration fit described in figure \ref{fig:BOB_calib}, which comes from data collected for a separate niobium sample plate, is included for reference.  The $1.6\,\text{K}$ data is that shown in figure \ref{fig:Q0_vs_B} and was limited by quench.  The maximum fields at $2.0\,\text{K}$ were not driven to quench.}
    \label{fig:R_vs_B}
\end{figure*}

\section{Effects of increasing gold thickness}
\label{sec:results_thickness}

For the majority of the measurements in the study, the resistance on the sample plate was large enough to be reliably extracted using equation \ref{eqn:Sample_R_extraction}.  The systematic error discussed in the previous section becomes less important for samples with thicker gold layers and at higher temperatures.  This section examines how gold layer thickness affects the extracted surface resistance dependence on field amplitude and temperature.

\begin{figure*}[!htbp]
    \centering
    \includegraphics[width = \linewidth]{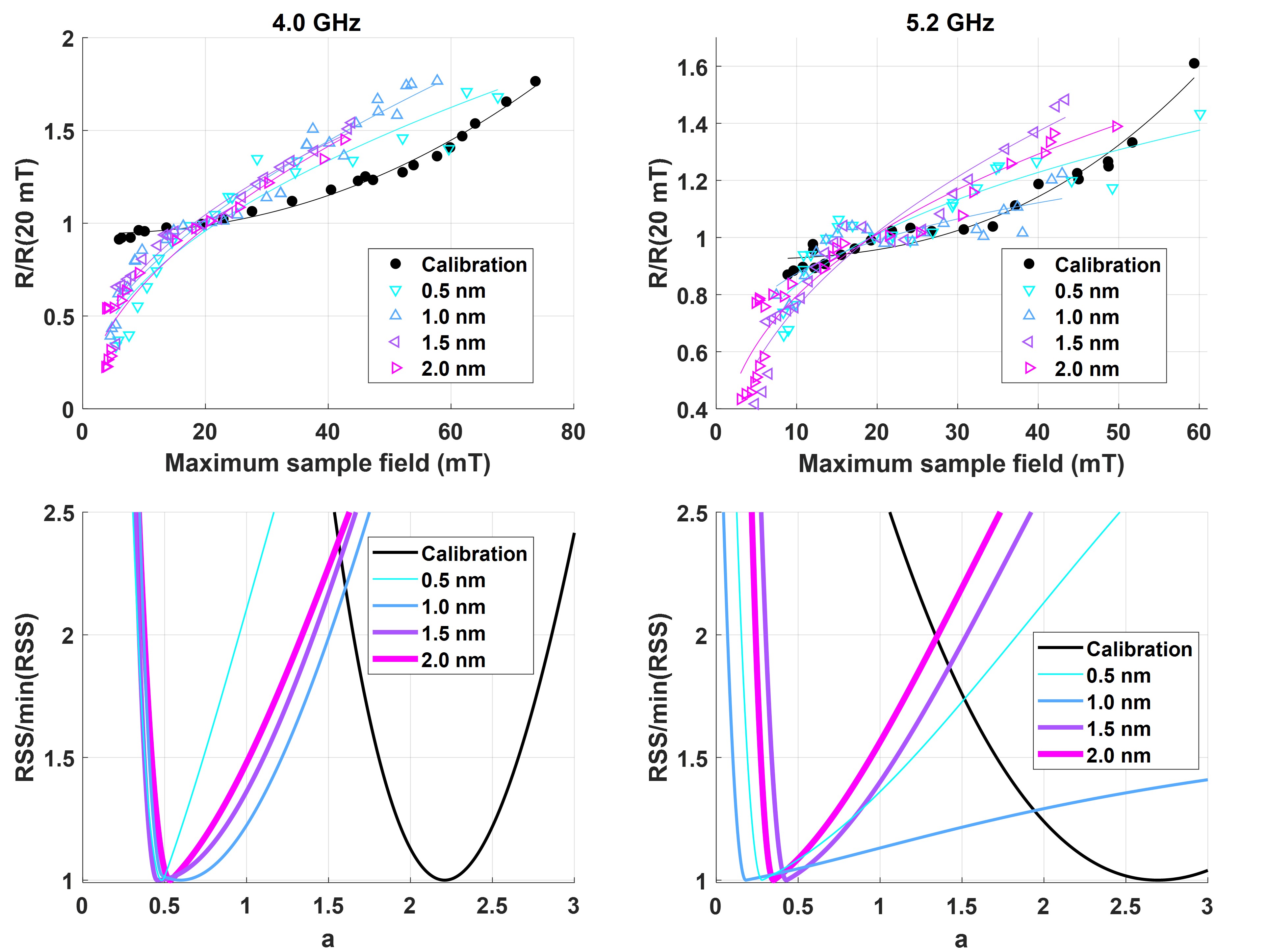}
    \caption{Extracted sample surface resistance (equation \ref{eqn:Sample_R_extraction}) from figure \ref{fig:R_vs_B} at $1.6\,\text{K}$ and $4.0\,\text{GHz}$ (top-left) and $5.2\,\text{GHz}$ (top-right) normalized by its value at $20\,\text{mT}$.  A power law, $R(B)/R(B_0) = \Lambda B^a + \Upsilon$, is fit to the normalized data requiring $\Lambda>0$ and $\Upsilon>0$ for a range of powers, $a$. $B$ is the RF field amplitude and $B_0$ is the field for normalization.  The resulting residual sum of squares (RSS) at $4.0\,\text{GHz}$ (bottom-left) and $5.2\,\text{GHz}$ (bottom-right) indicates the power which best fits the data.  The lines on the top plots indicate the best-fit power law (using $a$ which minimizes the RSS).}
    \label{fig:Normalized_R_vs_B}
\end{figure*}

\subsection{Field-dependence of the extracted surface resistance}


Applying equation \ref{eqn:Sample_R_extraction} to the measured quality factors in figure \ref{fig:Q0_vs_B} produces the sample surface resistances in figure \ref{fig:R_vs_B}.  The results are also included at $2.0\,\text{K}$ to convey the weak dependence of the surface resistance on temperatures in this range.  The surface resistance increases rapidly with gold thickness until the step from $1.5\,\text{nm}$ to $2.0\,\text{nm}$.  Here, the surface resistance increases only slightly.  This increase is more pronounced at $5.2\,\text{GHz}$.  Systematic error could be relevant for this behavior, but it is not expected to be the primary cause because of the relatively high sample resistance compared to the calibration.  The effective conductivity of the gold film, which is expected to increase with thickness~\cite{Gold_film_conductivity_meas_paper, Lacy2011}, could be relevant~\cite{Oseroff_thesis}. 

The surface resistance of the bilayer displays a strong increase at low fields and then becomes more gradual as the field increases further.  The field at which the qualitative change occurs decreases as gold thickness increases.  For the $1.5\,\text{nm}$ and $2.0\,\text{nm}$ thicknesses, the shift was extremely abrupt, which made exact measurement of the quality factor challenging.  The finer details of the measurements reported in figure \ref{fig:R_vs_B}, such as the behavior between $15\,\text{mT}$ and $25\,\text{mT}$ at $5.2\,\text{GHz}$, should be viewed with caution.  This is possibly an artifact of systematic effects~\cite{Oseroff_thesis}.  The focus of the reader should be on the broad dependence; that is the increase of the surface resistance with field and its overall form.  

To study the effects of gold film thickness on surface resistance field-dependence, the results of figure \ref{fig:R_vs_B} are normalized for each thickness to their values at a specific field.  Figure \ref{fig:Normalized_R_vs_B} shows the surface resistances normalized using the resistances at a field of $20\,\text{mT}$.  Similar results can be obtained by normalizing at other fields.  From this analysis, the qualitative field-dependence of the surface resistance appears roughly independent from thickness in the range explored.


\begin{figure*}[!htbp]
    \centering
    \includegraphics[width = \linewidth]{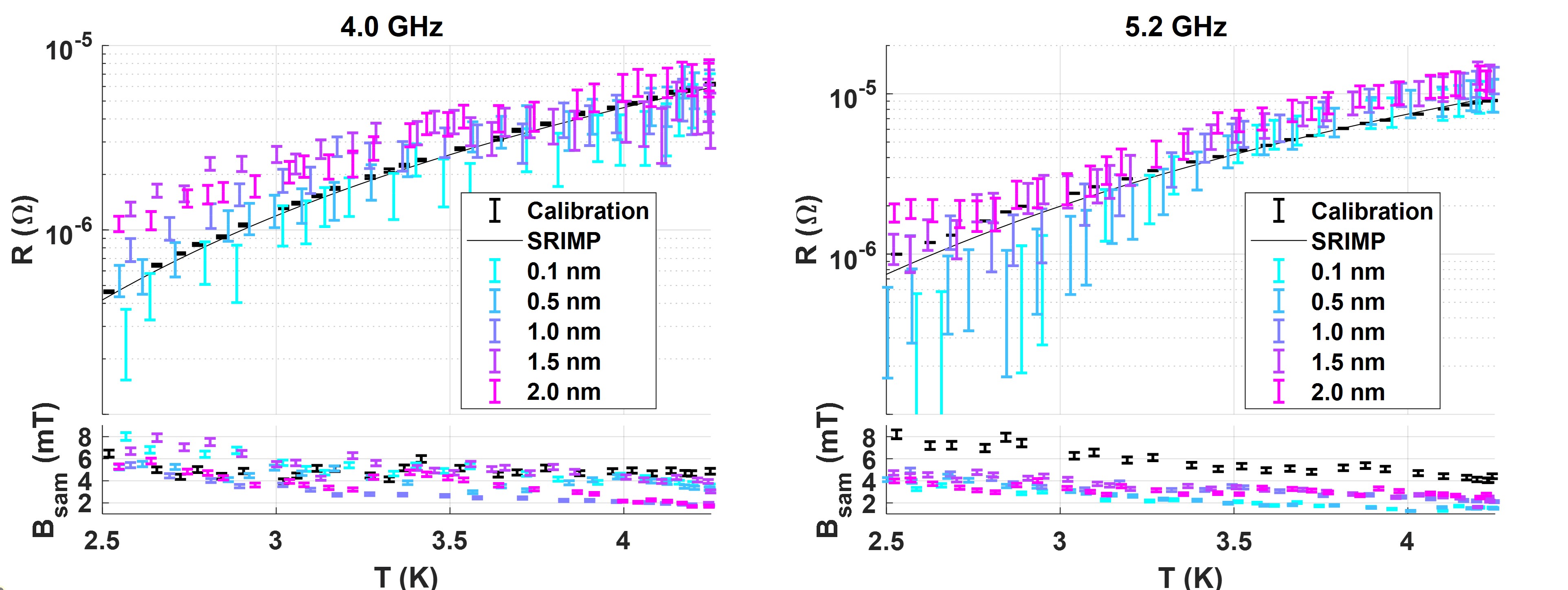}
    \caption{Extracted sample surface resistance (equation \ref{eqn:Sample_R_extraction}) of gold layers deposited over bulk niobium as described in section \ref{sec:sample_prep} plotted as a function of temperature at $4.0\,\text{GHz}$ (left) and $5.2\,\text{GHz}$ (right).  The RF magnetic field amplitude at which each measured resistance was obtained is presented below the surface resistance data.  The niobium calibration and its fit using the SRIMP model~\cite{halbritter_SRIMP}, displayed earlier in figure \ref{fig:BOB_calib}, are re-plotted here for reference.}
    \label{fig:RvsT}
\end{figure*}

The normalized field-dependent surface resistance in figure \ref{fig:Normalized_R_vs_B} is fit to a power law, $R(B)/R(B_0) = \Lambda B^a + \Upsilon$, with $\Lambda > 0$, $\Upsilon > 0$, and $a$ specified.  $B$ is the RF field amplitude and $B_0$ is the field for normalization.  The quality of the fit for a given value of $a$ is monitored with the residual sum of squares (RSS) as demonstrated in figure \ref{fig:Normalized_R_vs_B}.  At $4.0\,\text{GHz}$, all Au/Nb bilayers are best fit using a square root power law, $a = 0.5$.  For $5.2\,\text{GHz}$, there is more variation, but all bilayers are at a similar power.  The field-dependence using the value of $a$ which produces the best fit is included with the normalized data.  It is evident that a power law fits the $4.0\,\text{GHz}$ data reasonably well, but the agreement is worse at $5.2\,\text{GHz}$.  The discrepancy at $5.2\,\text{GHz}$ could be due to the previously mentioned features observed between $15\,\text{mT}$ to $25\,\text{mT}$.

Figure \ref{fig:Normalized_R_vs_B} also demonstrates that the field-dependence of the calibration niobium is clearly different than that of the Au/Nb bilayers.  Assuming the anticipated normal conducting phases in the native oxide are responsible for this change in behavior, it can be concluded that the normal conductor properties of such bilayers are important for determining the surface resistance field-dependence.  Thus, the observed independence of field-dependence on gold thickness in this study may be specific to the realized sample properties.  



\subsection{Temperature-dependence of the extracted surface resistance}

The surface resistance was measured with low power while slowly lowering the temperature at a similar rate to that described in section \ref{sec:BOB_cavity}. 
 As demonstrated in figure \ref{fig:R_vs_B}, the resistance variation at low temperatures was limited.  The following discussion will therefore be restricted to higher temperatures.  Here the bulk niobium, which has a characteristic temperature-dependence, is the dominant contribution.  Figure \ref{fig:RvsT} shows the results of the low-field measurements in this higher-temperature range.  This includes measurements of the Au/Nb bilayers, the calibration Nb measurement, and the fit of the calibration Nb to the surface impedance model, "SRIMP" described in section \ref{sec:BOB_cavity}.  
 For the temperatures considered in figure \ref{fig:RvsT}, the resistance of the Au/Nb bilayer with $0.1\,\text{nm}$ gold thickness can be resolved adequately and is included.

 At the highest temperatures measured, the surface resistance measurements are similar for all gold thicknesses.  This is expected, due to the high resistance of the substrate in this range.  As the temperature is reduced, the effect of the gold layer becomes visible.  For the lower temperatures in figure \ref{fig:RvsT} at $5.2\,\text{GHz}$, the resistance is seen to be increased for thicker gold layers.  At $4.0\,\text{GHz}$, the surface resistance of the bilayer with $1.5\,\text{nm}$ gold thickness appears slightly higher than that of the $2.0\,\text{nm}$ measurement.  This is due to the former measurements being performed at higher fields than the latter, and the strong field-dependence reported in figure \ref{fig:R_vs_B}.  The measurements in figure \ref{fig:RvsT} are subject to strong fractional uncertainties, but the surface resistance dependence on gold layer thickness is consistent with intuition and figure \ref{fig:R_vs_B}.


\section{Summary}

In this work, an attempt was made to measure the surface resistance of a bilayer structure, consisting of a thin normal conducting layer and a bulk superconductor, for a range of normal conductor thicknesses.  A sample was prepared by using acid to remove the native oxide and then transferring it to a gold evaporation chamber while attempting to minimize oxidation.  RF measurements were made using a sample host cavity capable of driving the samples to the maximum RF fields for which superconducting properties could be maintained (quench field).  Both the sample preparation and the RF measurements are imperfect and should be viewed with some caution.  Despite this, the results are reasonable and can be readily explained.


The surface resistance of the Au/Nb bilayer structure increases with gold thickness.  It appeared to scale with the square root of the applied RF field magnitude.  This field dependence seems to be independent from gold thickness, but was clearly different from that of a the untreated surface.

Attempting to replace the native oxide of a sample with a very thin gold layer appears to reduce its surface resistance.  The quench field was also enhanced, but it is likely this was due to the reduction in surface resistance.  The magnitude of the reduction could not be resolved, but was inferred by the change in measured quality factor.  
Because of systematic error, it cannot be concluded that the observed increase in quality factor of the thin gold layer measurement indicates an improvement over oxidized niobium in general.  The results only indicate that something in the procedure of section \ref{sec:sample_prep}, likely to do with the oxide, caused a reduction in the surface resistance of the sample plate.


\section{Outlook}

While it cannot be concluded that replacing the native niobium oxide with a thin gold layer reduces the surface resistance in general, it does appear that the passivation procedure improved the sample.  This implies that, at least in some cases, eliminating niobium oxide from RF surfaces could reduce surface resistance.  If validated, this result has utility for application.  

An important next step would be conducting measurements on a standard accelerator cavity with its surface passivated by a thin gold layer.  This would not suffer from the limitations of the calibrated quality factor measurement used in this work, and would probe behavior at lower frequencies used for accelerator applications.  At these frequencies, it is unlikely the improvement to the quench field would be observed.  The oxides of other materials considered for superconducting accelerator cavities, such as $\text{Nb}_3\text{Sn}$ and NbN, should also be considered as potential sources of spurious loss.

The primary focus of this work was on high-field superconducting radio frequency (SRF) cavity applications.  However, the passivation procedure may be relevant for applications where cavities operate at low fields and low temperatures.  Here, the limiting source of dissipation is thought to be two-level systems in the insulating portion of the niobium oxide~\cite{FNAL_TLS_2017}.  Some success has been achieved by reducing the thickness of this oxide, but the effects are still present~\cite{FNAL_TLS_2020}.  We suggest that a thin proximity-coupled gold film could be a solution to this problem.  The oxide, and the two-level systems within, could be entirely eliminated.  The surface resistance, observed here to be smaller or similar to the oxidized niobium for high amplitude fields, suggests a similar effect could be expected in the low-field and low-temperature limit.

\section*{Data Availability Statement}

The data that support the findings of this study are available
upon reasonable request from the authors.

\section*{Conflicts of Interest}

The authors declare no competing financial interests.

\section*{Acknowledgements}

This work was supported by the U.S. National Science Foundation under Award PHY-1549132, the Center for Bright Beams (CBB).  Zeming Sun assisted in picking up the choice of gold as a normal conducting and passivation layer on niobium in 2019, and extended the idea to the utilization of gold electroplating on the niobium SRF cavities and provided the proof-of-concept demonstrations.  Another member of the CBB, Nathan Sitaraman, proposed introducing substitutional gold impurities to a niobium surface as a method of suppressing oxide to improve cavity performance in 2021.  This work made use of the Cornell Center for Materials Research Shared Facilities which are supported through the NSF MRSEC program (DMR-1719875), and was performed in part at the Cornell NanoScale Facility, an NNCI member supported by NSF Grant NNCI-2025233.

\section*{References}

\bibliographystyle{unsrt}
\bibliography{references.bib}

\end{document}